\RequirePackage[displaymath]{lineno}

\documentclass[twocolumn,aps,prl,showpacs,superscriptaddress,floatfix]{revtex4}
\usepackage{newtxtext,newtxmath,booktabs,siunitx}
\usepackage{dcolumn}
\usepackage{bm}
\usepackage{float}
\usepackage[]{graphicx}  
\usepackage{booktabs}
\usepackage{tabularx}
\usepackage{amsmath,bm}
\usepackage{xcolor}
\usepackage{multirow}
\usepackage[colorlinks,citecolor=blue,urlcolor=blue,linkcolor=blue]{hyperref}

\newcommand {\Npart}	{N_{\rm part}}
\newcommand {\Zr}	{$^{96}$Zr}
\newcommand {\Ru}	{$^{96}$Ru}
\newcommand {\RuRu}	{$^{96}_{44}$Ru+$^{96}_{44}$Ru}
\newcommand {\ZrZr}	{$^{96}_{40}$Zr+$^{96}_{40}$Zr}
\newcommand {\RuZr}	{Ru+Ru/Zr+Zr}
\newcommand {\Pb}	{$^{208}$Pb}
\newcommand {\rnp}	{$\Delta r_{\rm np}$}
\newcommand {\Lc}	{L(\rho_{c})}

\newcommand {\pt}	{p_{\perp}}
\newcommand {\meanpT}	{\langle p_{\perp}\rangle}
\newcommand {\Rpt}	{R_{\meanpT}}
\newcommand {\Rdt}	{R_{\dt}}
\newcommand {\St}	{S_{\perp}}
\newcommand {\rt}	{r_{\perp}}
\newcommand {\dt}	{d_{\perp}}
\newcommand {\ecc}	{\epsilon_2}

\newcommand {\mean}[1]	{\langle #1\rangle}

\begin{document}
\title{Probing nuclear structure with mean transverse momentum in relativistic isobar collisions}
\author{Hao-jie Xu}
\affiliation{School of Science, Huzhou University, Huzhou, Zhejiang 313000, China}
\author{Wenbin Zhao}
\affiliation{Department of Physics and Astronomy, Wayne State University, Detroit, Michigan 48201, USA}
\author{Hanlin Li}
\affiliation{College of Science, Wuhan University of Science and Technology, Wuhan, Hubei 430065, China}
\author{Ying Zhou}
\affiliation{
School of Physics and Astronomy, Shanghai Key Laboratory for Particle Physics and Cosmology, and Key Laboratory for Particle Astrophysics and Cosmology (MOE), Shanghai Jiao Tong University, Shanghai 200240, China
}
\author{Lie-Wen Chen}
\affiliation{
School of Physics and Astronomy, Shanghai Key Laboratory for Particle Physics and Cosmology, and Key Laboratory for Particle Astrophysics and Cosmology (MOE), Shanghai Jiao Tong University, Shanghai 200240, China
}
\author{Fuqiang Wang}
\affiliation{School of Science, Huzhou University, Huzhou, Zhejiang 313000, China}
\affiliation{Department of Physics and Astronomy, Purdue University, West Lafayette, Indiana 47907, USA}

\date{\today}
\begin{abstract}
Transverse momentum ($\pt$) generation in relativistic heavy ion collisions is sensitive to the initial geometry and the final-state bulk evolution. We demonstrate with hydrodynamic calculations that the mean $\pt$ ratio ($\Rpt$) between the highly similar isobar \RuRu\ and \ZrZr\ collisions is insensitive to the bulk evolution and remains sensitive to the small difference in the initial nuclear structure (neutron skin and  deformation) between the Ru and Zr nuclei. 
We further find that nuclear deformation can produce an anticorrelation between $\Rpt$ and eccentricity (or elliptic flow) in central collisions. These findings suggest that the $\Rpt$ between the isobar systems can be used to measure the neutron skin thickness and deformation parameters, which can in turn constrain the nuclear symmetry energy slope parameter.
\end{abstract}

\maketitle


{\em Introduction.}
Relativistic heavy ion collisions at BNL's Relativistic Heavy Ion Collider (RHIC) and CERN's Large Hadron Collider (LHC) create a strongly coupled quark-gluob plasma (QGP), governed by quantum chromodynamics (QCD)~\cite{Adams:2005dq,Adcox:2004mh,ALICE:2010suc,Gyulassy:2003mc}. The evolution of the QGP medium can be successfully described by relativistic hydrodynamics with a nearly minimum value of shear viscosity to entropy density ratio $\eta/s$ 
in nature~\cite{Gale:2013da,Kovtun:2004de,Romatschke:2007mq,Song:2010mg,Xu:2016hmp,Zhao:2017yhj}.
The mean transverse momentum ($\meanpT$) of hadrons in relativistic heavy ion collisions reflects the expansion strength of the formed hot and dense QCD medium. 
With the same total entropy (energy), 
a denser initial condition would lead to a faster expansion, and thus a lager radial flow and $\meanpT$~\cite{Broniowski:2009fm,Bozek:2012fw,Mazeliauskas:2015efa,Schenke:2020uqq,Giacalone:2020dln}. 
This connection between initial collision geometry and final observables provides a potential opportunity to probe the structure of the colliding nuclei~\cite{Giacalone:2019pca,Jia:2021qyu}. 
However, the magnitude of $\meanpT$ itself strongly depends on the bulk properties of the medium--to describe experimental data, 
a finite bulk  viscosity is required in hydrodynamic calculations~\cite{Ryu:2015vwa,Bernhard:2019bmu}. Thus, in order to probe the initial geometry, 
one has to rely on $\meanpT$ fluctuations and correlations to  anisotropic flow 
 to eliminate  large uncertainties in the $\meanpT$ magnitudes caused by uncertainties in the dynamic evolution~\cite{Broniowski:2009fm,Bozek:2012fw,Mazeliauskas:2015efa,Schenke:2020uqq,Giacalone:2020dln}. 
 
 On the other hand, one may exploit collisions of isobar nuclei where the dynamic evolution is similar--so its uncertainties cancel in their comparisons--but the initial geometries are different.
 The isobar collisions of \RuRu\ and \ZrZr\ were originally proposed to control the background in  search for the chiral magnetic effect (CME)~\cite{Kharzeev:2007jp,Voloshin:2010ut,Kharzeev:2015znc,Deng:2016knn,Zhao:2019hta,STAR:2019bjg}. Such collisions were conducted in 2018, and the STAR collaboration has collected $\sim$2 billion events for each collision species~\cite{STAR:2021mii}. 
 Previous studies indicate that the nuclear density distributions of the two isobar nuclei differ~\cite{Xu:2017zcn}. The nuclear structure difference can cause significant observable differences between the isobar systems, such as in their event multiplicities  and elliptic flows that are  crucial to the CME search~\cite{Xu:2017zcn,Li:2018oec,Xu:2021vpn}.  
 Indeed, those differences have been observed in isobar data~\cite{STAR:2021mii} and are consistent with the predictions 
 from energy density functional theory (DFT) calculations~\cite{Xu:2017zcn,Li:2018oec,Xu:2021vpn}. The DFT calculations indicate a large halo-type neutron skin thickness (\rnp) for the \Zr\ nucleus~\cite{Li:2019kkh}; the \rnp\ is $0.183$~fm for \Zr\ and $0.042$~fm for \Ru\ with a reasonable parameter set (see~\cite{Li:2019kkh}). 
The neutron skin difference between \Zr\ and \Ru\ comes from the neutron-proton asymmetry in nuclear matter equation of state (EOS), which is encoded by the symmetry energy~\cite{Chen:2005ti,RocaMaza:2011pm,Tsang:2012se}. 

The DFT calculations are based on well-established nucleon-nucleon potential parameters in the extend Skyrme-Hartree-Fock model fitting experimental data~\cite{Tamii:2011pv,Zhang:2014yfa}. 
The density slope parameter of the symmetry energy is fitted to be $\Lc=47.3$~MeV at a subsatruration cross density $\rho_{c}=0.11\rho_{0}/0.16\simeq0.11$~${\rm fm}^{-3}$~\cite{Zhang:2013wna,Zhang:2014yfa,Zhou:2019omw}.
The DFT calculation with the same model parameters give a neutron skin \rnp\ = $0.190$~fm for the benchmark \Pb\ nucleus~\cite{Li:2019kkh}. 
However, the recent PREX-II measurement using parity-violating electroweak interactions has yielded a large neutron skin thickness of the \Pb\ nucleus \rnp\ = $0.283\pm0.071$~fm~\cite{Adhikari:2021phr}, 
leading to a larger density slope parameter $\Lc=71.5 \pm 22.6$~MeV~\cite{Reed:2021nqk}, at tension with the world data established by strong interaction means.

Owing to the large statistics of isobar collisions, the differences between two collision systems can be measured very precisely. 
We have proposed that the multiplicity difference between the isobar collisions can be used to probe the neutron skin and the symmetry energy slope parameter~\cite{Li:2019kkh}. Some of us have suggested that the elliptic flow measurements can also determine the proper nuclear structures of the isobar nuclei~\cite{Xu:2021vpn}.
In this paper, we will further show that the $\meanpT$ ratio in relativistic isobar collisions, 
\begin{equation}
	\Rpt = \frac{\meanpT^{\rm Ru+Ru}} {\meanpT^{\rm Zr+Zr}}\,,
	\label{eq:ratio}
\end{equation}
can also be used to probe nuclear structures,
and this ratio has weak dependence on transport properties of the collision system.
We also study the effect of nuclear deformation on $\Rpt$, and demonstrate, together with elliptic flow measurements, that both the neutron skin thickness and the deformation can be determined.
This work supplements our previous work on multiplicity difference between the isobar systems in probing nuclear structures.

{\em Model description and initial conditions.}
In this study, the $\meanpT$ is 
calculated by the iEBE-VISHNU model~\cite{Song:2007ux,Shen:2014vra,Bernhard:2016tnd}, an event-by-event (2+1)-dimensional viscous hydrodynamics, together with the hadron cascade model (UrQMD) to simulate the evolution of the subsequent hadronic matter~\cite{Bass:1998ca,Bleicher:1999xi}.
The initial condition of the collision is  obtained by the Trento model~\cite{Moreland:2014oya,Bernhard:2016tnd}, given   a nuclear density ditribution.
After a short  free streaming, 
the evolution of the initial energy momentum tensor in hydrodynamics follows the conservation law $\partial_{\mu}T^{\mu\nu}=0$. All the parameters for the iEBE-VISHNU simulation are taken from ~\cite{Bernhard:2019bmu} 
which were calibrated to experimental data at the LHC, 
except the normalization factor to match the multiplicity in isobar collisions at RHIC~\cite{STAR:2021mii}.

For the initial condition, the nuclear density distribution of the colliding nuclei is incorporated into the Trento model. 
Similar to our previous study~\cite{Li:2019kkh,Xu:2021vpn,Xu:2021qjw}, the isobar nuclear densities are assumed to be spherical and calculated by DFT with three density slope parameters, $\Lc=20, 47.3, 70$ MeV. The calculated densities are parameterized~\cite{Xu:2021vpn} by the Woods-Saxon (WS) distributions
\begin{eqnarray}
	\rho &=& \frac{\rho_{0}}{1+\exp{(r-R)/a}}\,,\\
	R &=& R_{0}\left(1 + \beta_{2}Y_{2}^{0} + \beta_{3}Y_{3}^{0}+\cdots\right)\,,
\end{eqnarray}
where the deformity parameters $\beta_2$ and $\beta_3$ are set to zero.
The  $R_0$ and $a$ parameters are determined by matching the $\mean{r}$ and $\mean{r^2}$ quantities from the DFT-calculated densities~\cite{Xu:2021vpn}.
The $\rho_0$ parameter is fixed by normalization of the nucleus volume.
The corresponding WS parameters are listed in Table~\ref{tab:WSDFT}. 
Since we use the WS parameterizations instead of the DFT-calculated densities throughout the  paper, 
we simply denote those WS densities as Lc20, Lc47,and Lc70. 
\begin{table}
	\caption{WS parameterizations (radius parameter $R_0$ and diffuseness parameter $a$) of the \Ru\ and \Zr\ nuclear density distributions, matching to the corresponding $\mean{r}$ and $\mean{r^2}$ from the Lc20, Lc47, and Lc70 spherical densities calculated by DFT. The $\rho_0$ is fixed by volume normalization.
	The WS parameterization of the \Ru\ (\Zr) nuclear density with an assumed quadrupole (octupole) deformity parameter $\beta_{2}=0.16$ ($\beta_{3}=0.16$), keeping the $\rho_0$ value and matching to the volume and RMS radius of the spherical Lc47 density, is also listed. The quoted values for $R_0$ and $a$ are in fm and that for $\rho_0$ is in 1/fm$^3$. \label{tab:WSDFT}}
	\centering{}%
    \begin{tabular}
	    {p{1.1cm}p{0.8cm}p{0.8cm}p{0.8cm}p{0.7cm}p{0.8cm}p{0.8cm}p{0.8cm}p{0.7cm}}
    \hline
     & \multicolumn{4}{c}{\Ru} & \multicolumn{4}{c}{\Zr}  \\
	       & $\rho_{0}$ & $R$    & $a$   & $\beta_2$ & $\rho_{0}$ & $R$ & $a$ & $\beta_3$ \\
    \hline
	    Lc20  & 0.161 & 5.076 & 0.483 & 0.00 & 0.166 & 4.994 & 0.528 & 0.00  \\
	    Lc47  & 0.159 & 5.093 & 0.488 & 0.00 & 0.163 & 5.022 & 0.538 & 0.00  \\
	    Lc70  & 0.157 & 5.114 & 0.487 & 0.00 & 0.160 & 5.045 & 0.543 & 0.00  \\
	\hline
	    Lc47Def   & 0.159 & 5.090 & 0.473 & 0.16 & 0.163 &5.016  & 0.527  & 0.16  \\
	\hline
	\end{tabular}
\end{table}

There is strong evidence from anisotropic flow ($v_n$) measurements in central isobar collisions that the Ru and Zr nuclei have different deformations~\cite{STAR:2021mii}.
Those measurements suggest~\cite{Zhang:2021kxj} that the \Ru\ nucleus has a quadrupole deformation,
while the \Zr\ nucleus has a octupole deformation.
DFT calculations of deformed nuclei are challenging and usually yield large uncertainties.
It is not clear how deformation affects the nucleus size with respect to the spherical one's. 
In this study, the WS parameters $R_0$ and $a$ for the given $\beta_2=0.16$ (or $\beta_3=0.16$) are calculated to match the volume and RMS of the corresponding nucleus calculated by DFT with $\Lc=47.3$ MeV, 
keeping the normalization $\rho_0$ fixed.
These parameters are also listed in Table~\ref{tab:WSDFT}, denoted as Lc47Def.

It is well known that the $\meanpT$ is related to the transverse energy density in heavy ion collisions~\cite{Broniowski:2009fm,Schenke:2020uqq}, approximately by
\begin{equation}
    \meanpT\propto\dt\equiv\sqrt{\Npart/\St}\,,
    \label{eq:meanpt}
\end{equation}
where $\Npart$ is the number of participant nucleons and $\St$ is the transverse overlap area. 
The neutron skin affects $\St$, so the $\meanpT$ is sensitive to the neutron skin thickness.
Because the latter depends on  $\Lc$, the $\meanpT$ can be used to probe the $\Lc$ parameter.
The overlap area can be calculated by~\cite{Bhalerao:2005mm,Schenke:2020uqq}
\begin{equation}
	\St = \pi\sqrt{\langle x^2\rangle\langle y^{2}\rangle - \langle xy\rangle^{2}}\equiv\pi\langle r_\perp^{2}\rangle\sqrt{1-\epsilon_{2}^{2}}\,,
	\label{eq:St}
\end{equation}
where
\begin{eqnarray}
	\langle r_{\perp}^{2}\rangle &\equiv& \langle x^{2} + y^{2}\rangle, \\ 
	\epsilon_{2}^{2} &\equiv& \frac{(\mean{y^{2}} - \mean{x^{2}})^2 + 4\langle xy \rangle^{2}}{\mean{x^{2}+y^{2}}^{2}} 
\end{eqnarray}
are the root-mean-square (RMS) and eccentricity of the overlap area, respectively. 
It is clear from Eq.~(\ref{eq:St}) that nuclear deformation also affects $\St$.
In the spherical case, $\rt$ is affected by the neutron skin; in the deformed case, $\rt$ is affected by both the neutron skin and the $\beta_2$.
The $\meanpT(\St)$ is centrality dependent and model dependent. These dependence, however, are largely canceled in the $\meanpT$ ratio in Eq.~(\ref{eq:ratio}).

In this study, we simulate $\sim$1.6M hydrodynamic events each for  Zr+Zr  and Ru+Ru collisions, 
together with $50$ oversampling of UrQMD afterburner for each hydrodynamic event. With such statistics, we are able to determine $\Rpt$ to the precision of $10^{-5}$ in a given centrality, so the statistical errors are not visible for the results presented in this paper. Even considering potential underestimation of the statistical uncertainties from the UrQMD oversampling, the precision is still well under control, especially for the most central collisions. 
We note that the $\Rpt$ can be measured even more precisely in experiment, where $\sim$2B good events have been collected for each isobar collision system~\cite{STAR:2021mii}.

\begin{figure} 
	\includegraphics[scale=0.22]{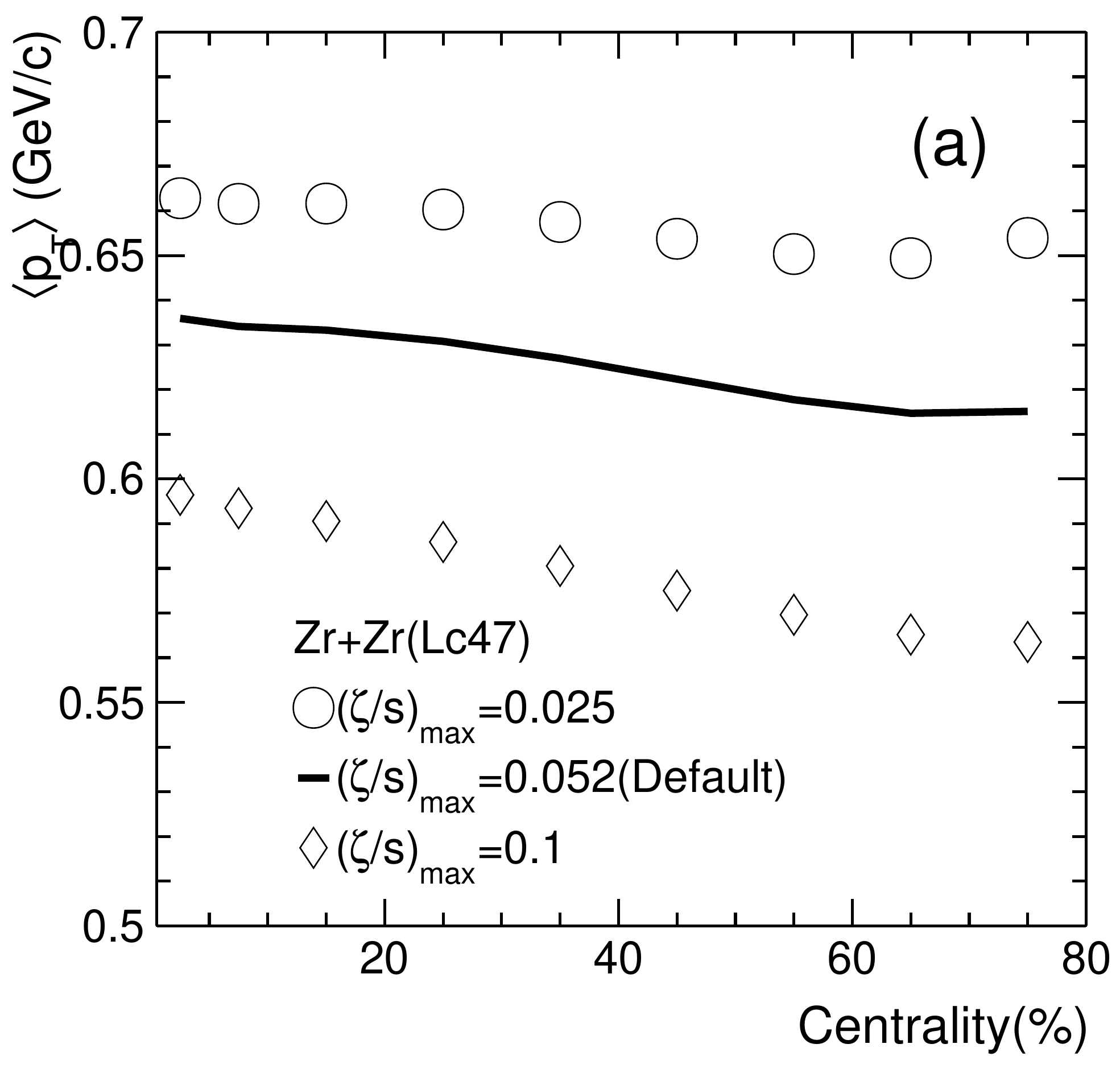}\includegraphics[scale=0.22]{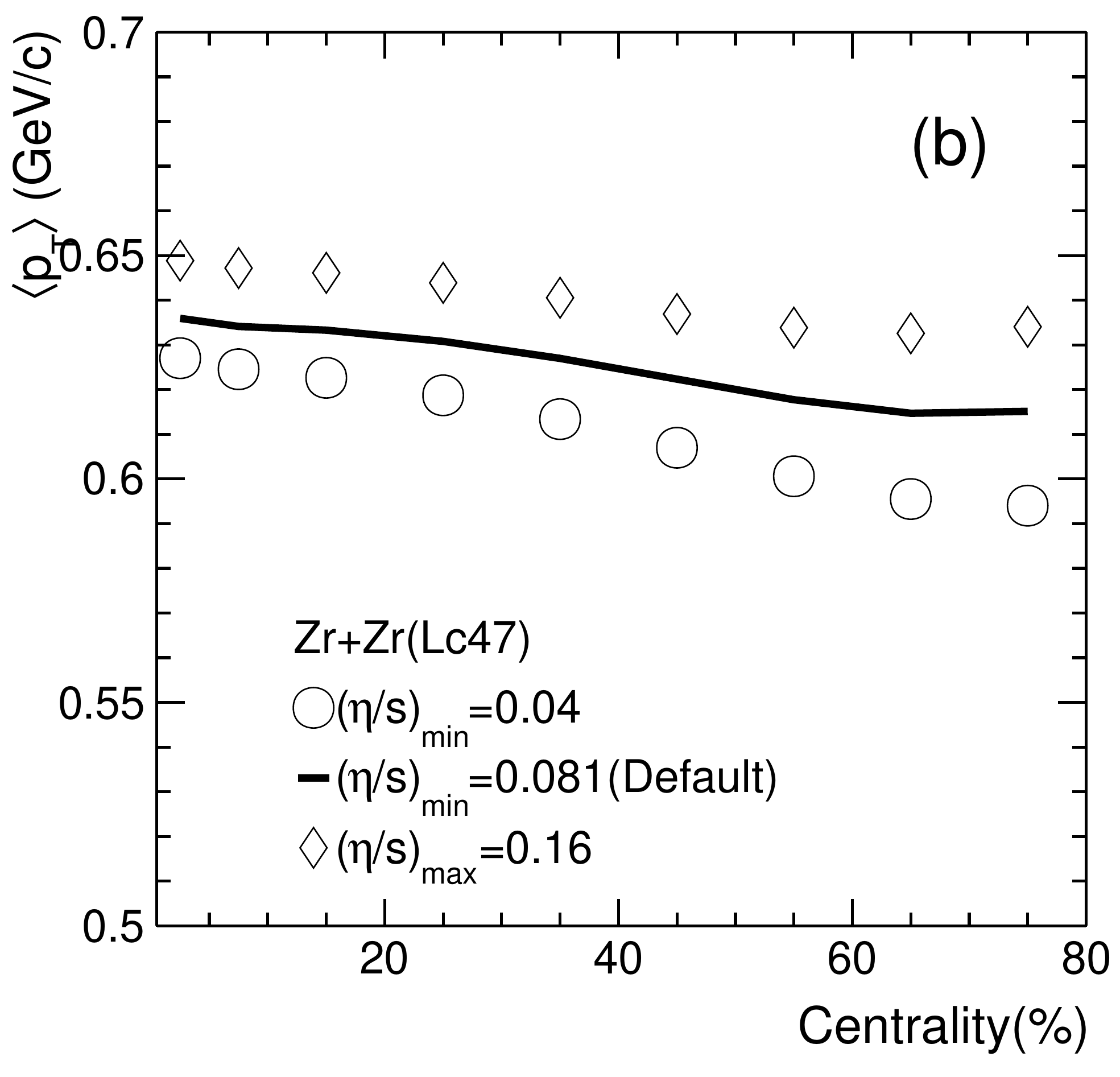}
	\includegraphics[scale=0.22]{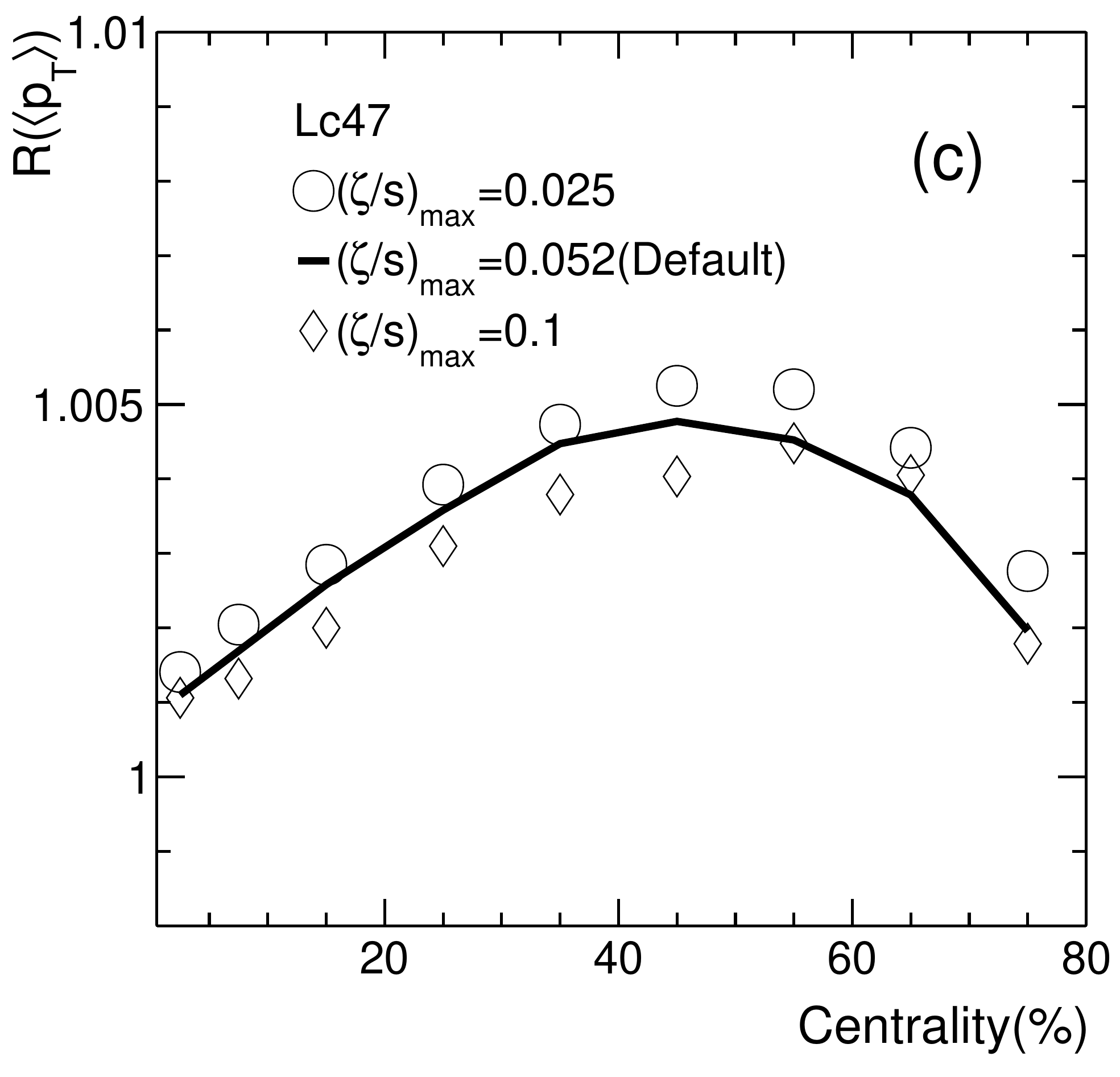}\includegraphics[scale=0.22]{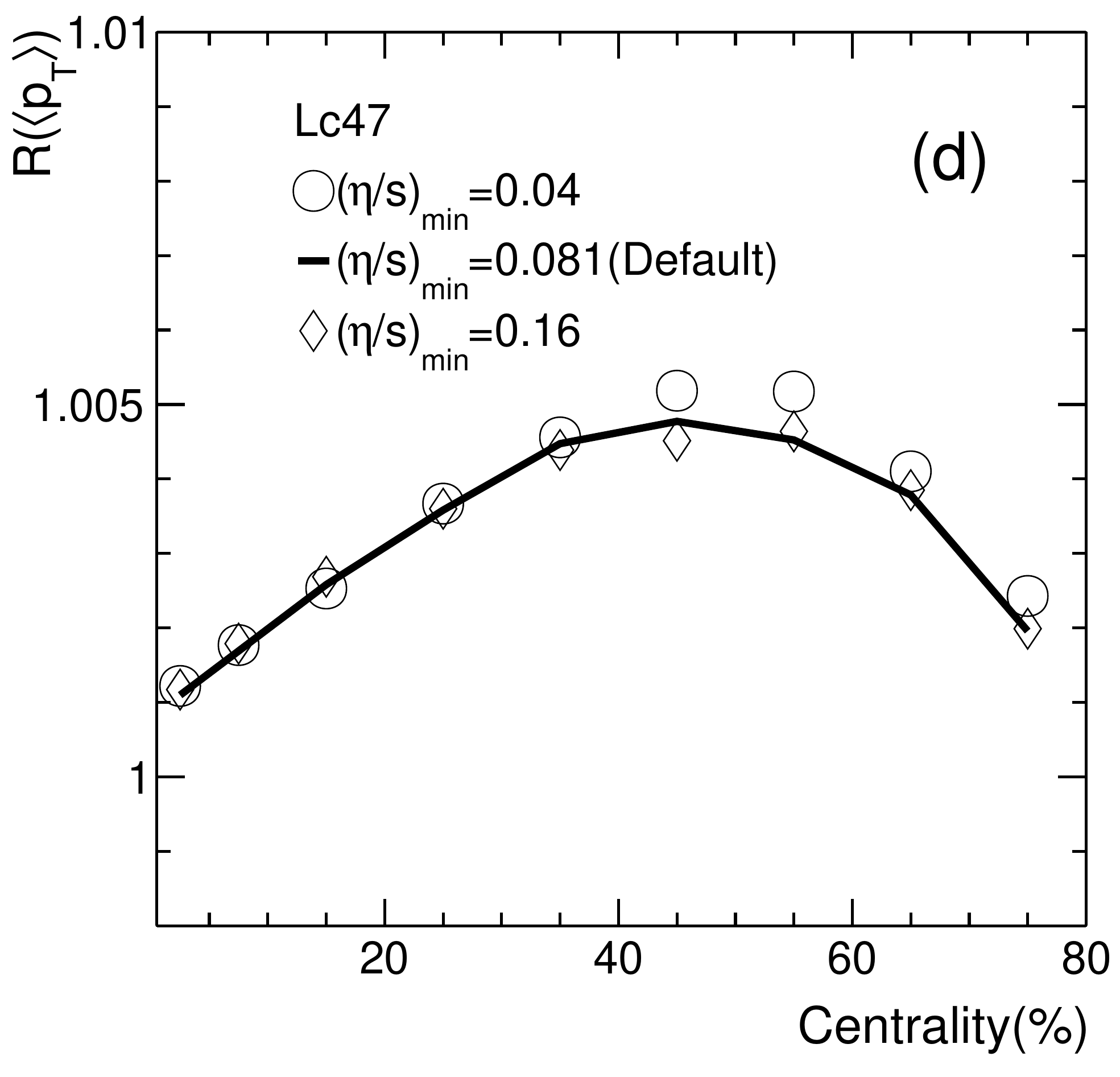}
      \caption{(a,b) The mean transverse momentum $\meanpT$ as functions of centrality in Zr+Zr collisions, calculated by the iEBE-VISHNU model with different $(\eta/s)_{\rm min}=0.04, 0.08, 0.12$ and $(\zeta/s)_{\rm max}=0.025, 0.052, 0.1$ with the Lc47 densities.  (c,d) The corresponding \RuZr\ ratio $\Rpt$.
      \label{fig:meanpTBS}} 
 \end{figure}

{\em Results.}
In hydrodynamics, the $\meanpT$ values are sensitive to the medium bulk properties.
To investigate the effects of bulk properties, we calculate the $\meanpT$ using the Lc47 densities with three values of shear viscosity ($(\eta/s)_{\rm min}=0.04, 0.08$ and $0.16$) and with three values of bulk viscosity  ($(\zeta/s)_{\rm max}=0.025, 0.081$ and $0.1$), respectively. The middle values are typical values  used in hydrodynamic simulations~\cite{Bernhard:2019bmu}.
The results are depicted in Fig.~\ref{fig:meanpTBS}(a,b), which show strong sensitivities of $\meanpT$ to those bulk properties, especially to the bulk viscosity,  consistent with
previous studies~\cite{Ryu:2015vwa}.
Those bulk properties, however, have little effect on the centrality-dependent ratio $\Rpt$  between Ru+Ru and Zr+Zr collisions, as shown Fig.~\ref{fig:meanpTBS}(c,d).
The largest variation in $\Rpt$ from the relatively large ranges of the medium viscosity is on the order of 0.001.
This finding strongly indicates, while the magnitude of $\meanpT$ depends on the bulk properties, the ratio $\Rpt$ is insensitive to them and hence their  uncertainties. 
\begin{figure*} 
      \includegraphics[scale=0.35]{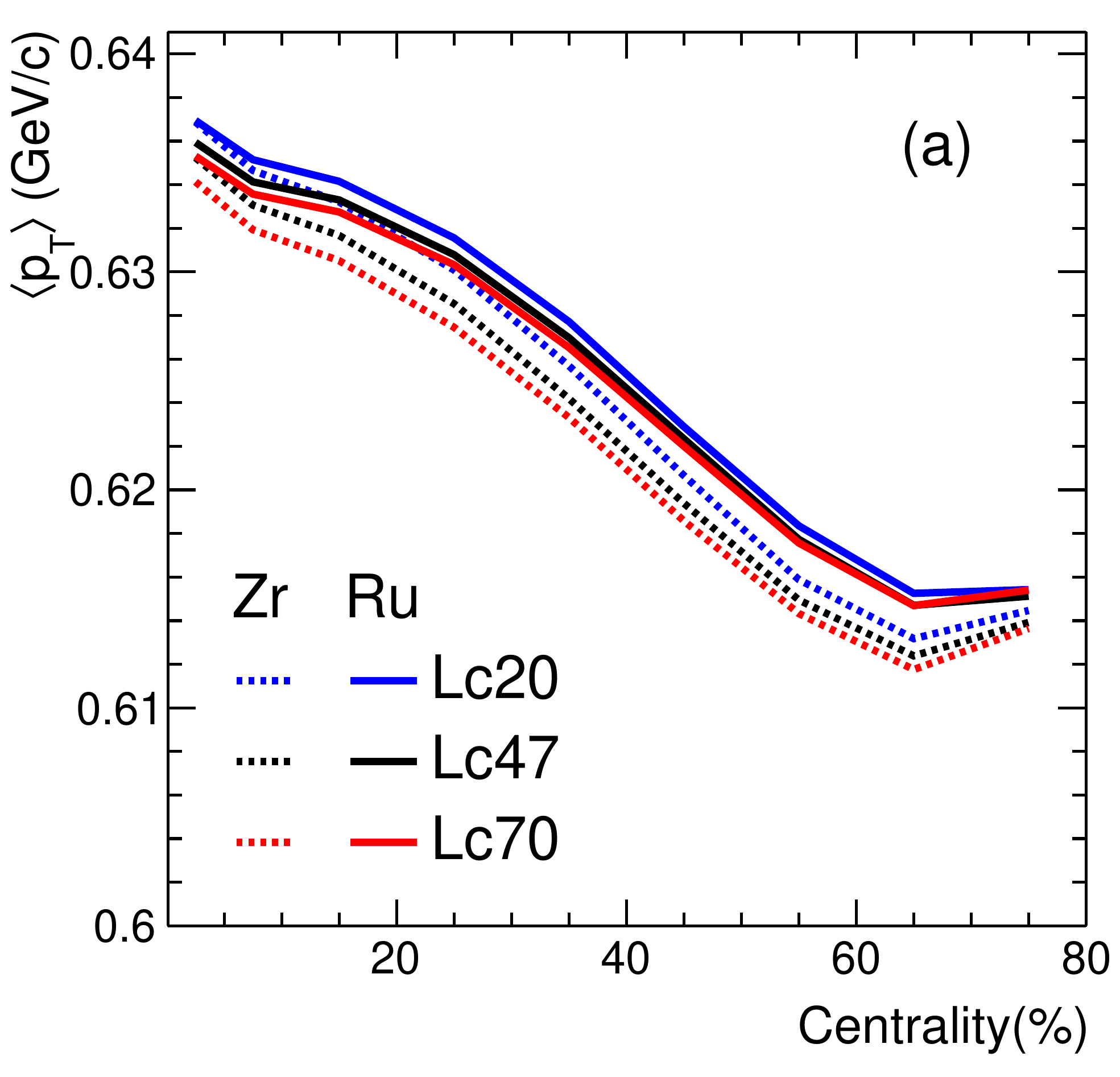}
	\includegraphics[scale=0.35]{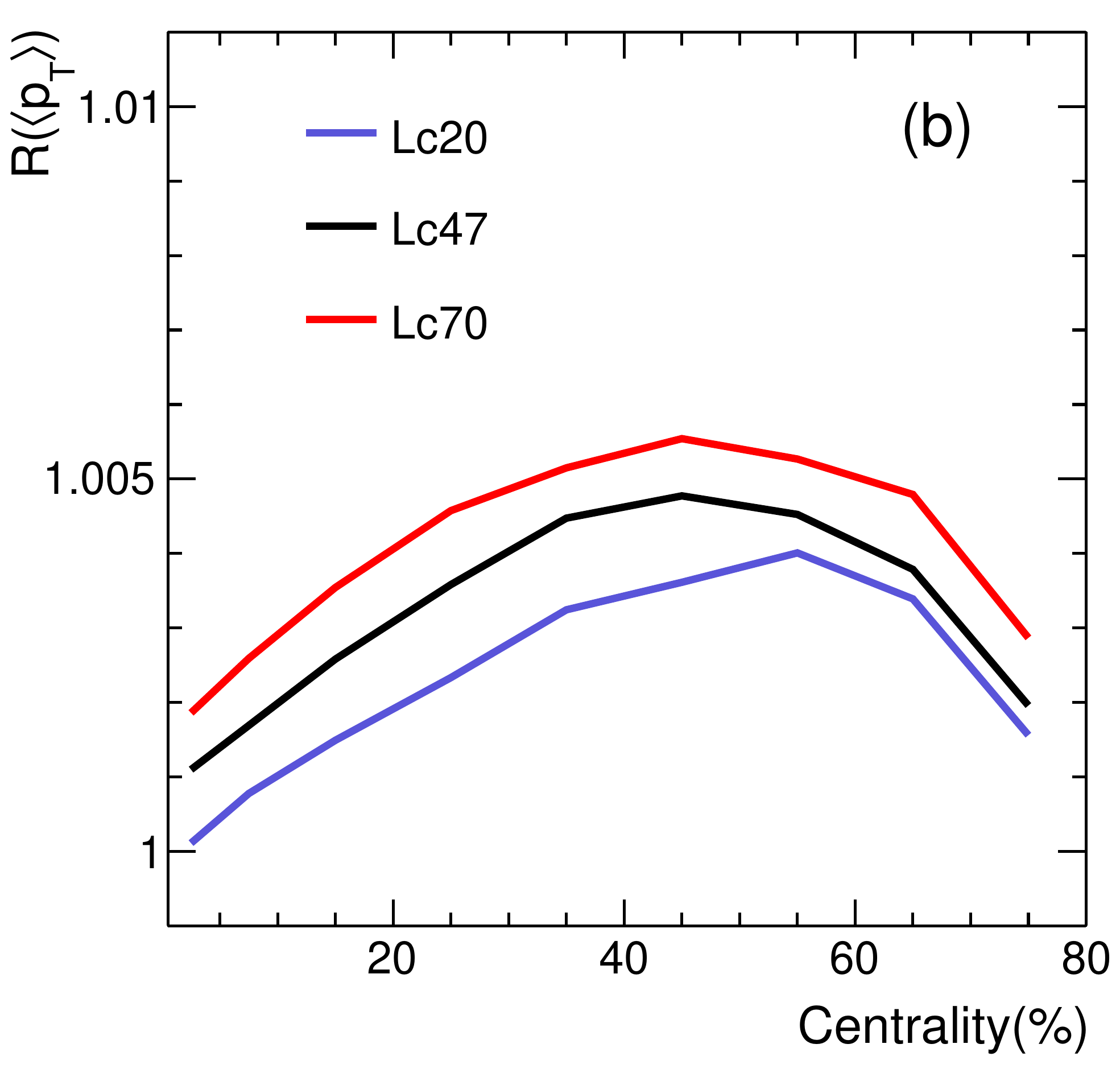}
      \caption{(Color online).
	  (a) The mean transverse momentum $\meanpT$ in Zr+Zr collisions, and (b) the \RuZr\ ratio $\Rpt$ as functions of centrality,  calculated by the iEBE-VISHNU model with Lc20, Lc47, and Lc70 spherical nuclear densities.
      \label{fig:meanpTA}} 
 \end{figure*}

Having demonstrated the insensitivity of $\Rpt$ to the the bulk properties, we now investigate effects of the initial condition of nuclear density. 
Figure~\ref{fig:meanpTA}(a) presents the $\meanpT$ as functions of centrality in both Ru+Ru  and Zr+Zr collisions from the iEBE-VISHNU simulations with various DFT-calculated sphetical densities for the isobars.
Larger $\Lc$ gives thicker neutron skin and larger $\St$, and results in smaller $\meanpT$ at each centrality, as expected. 
On the other hand, the the \RuZr\ ratio $\Rpt$, shown in Fig.~\ref{fig:meanpTA}(b), increases with $\Lc$. 
This is because the neutron skin effect is larger in \Zr\ than in \Ru\ and this effect increases with $\Lc$. 
The centrality dependence of $\Rpt$ is non-trivial, and can reach as large as 0.5\% above unity.

\begin{figure}
      \includegraphics[scale=0.35]{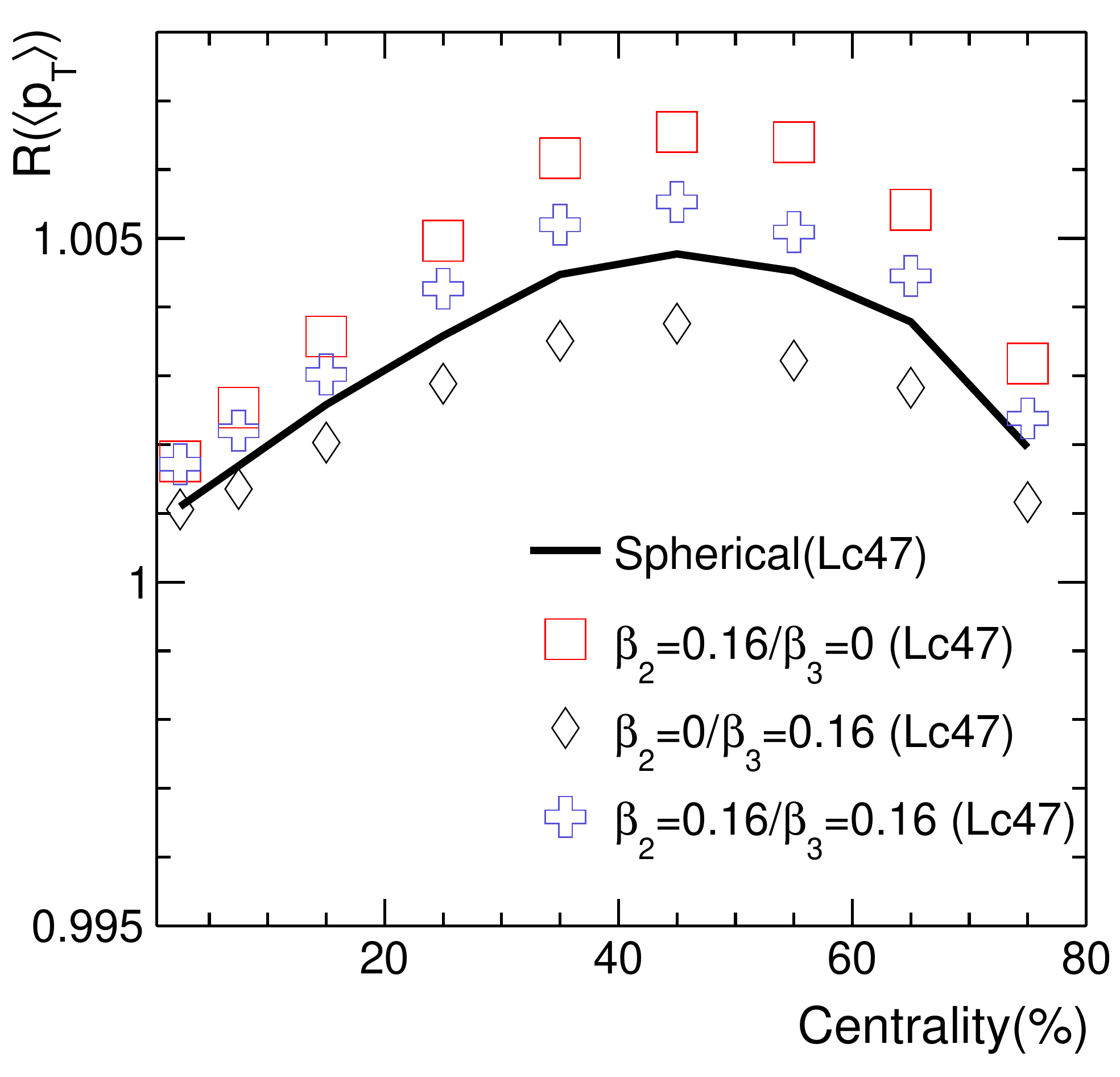}
      \caption{(Color online). The \RuZr\ ratio $\Rpt$ as functions of centrality, calculated by the iEBE-VISHNU model with the Lc47 nuclear densities and various deformation parameters for the \Ru\ and \Zr\ nuclei.
      \label{fig:Deformation}} 
 \end{figure}

Figure~\ref{fig:Deformation} shows $\Rpt$ as functions of centrality for various deformation differences between Ru and Zr. 
The finite quadrupole deformation $\beta_{2}=0.16$ for Ru gives a significant increase in $\Rpt$, and this increase is larger in non-central collisions. This is because $\beta_{2}$ effectively compresses the size of the overlap area (see Eq.~(\ref{eq:St})), generating larger $\meanpT$. 
A finite octupole deformation $\beta_{3}=0.16$ for Zr gives a negative effect to $\Rpt$ in non-central collisions. This is  because a finite $\beta_{3}$ can introduce an effective $\epsilon_{2}$~\cite{Jia:2021tzt}.
The effect of $\beta_3$ on $\Rpt$ is generally smaller than that of $\beta_2$.

\begin{figure} 
      \includegraphics[scale=0.35]{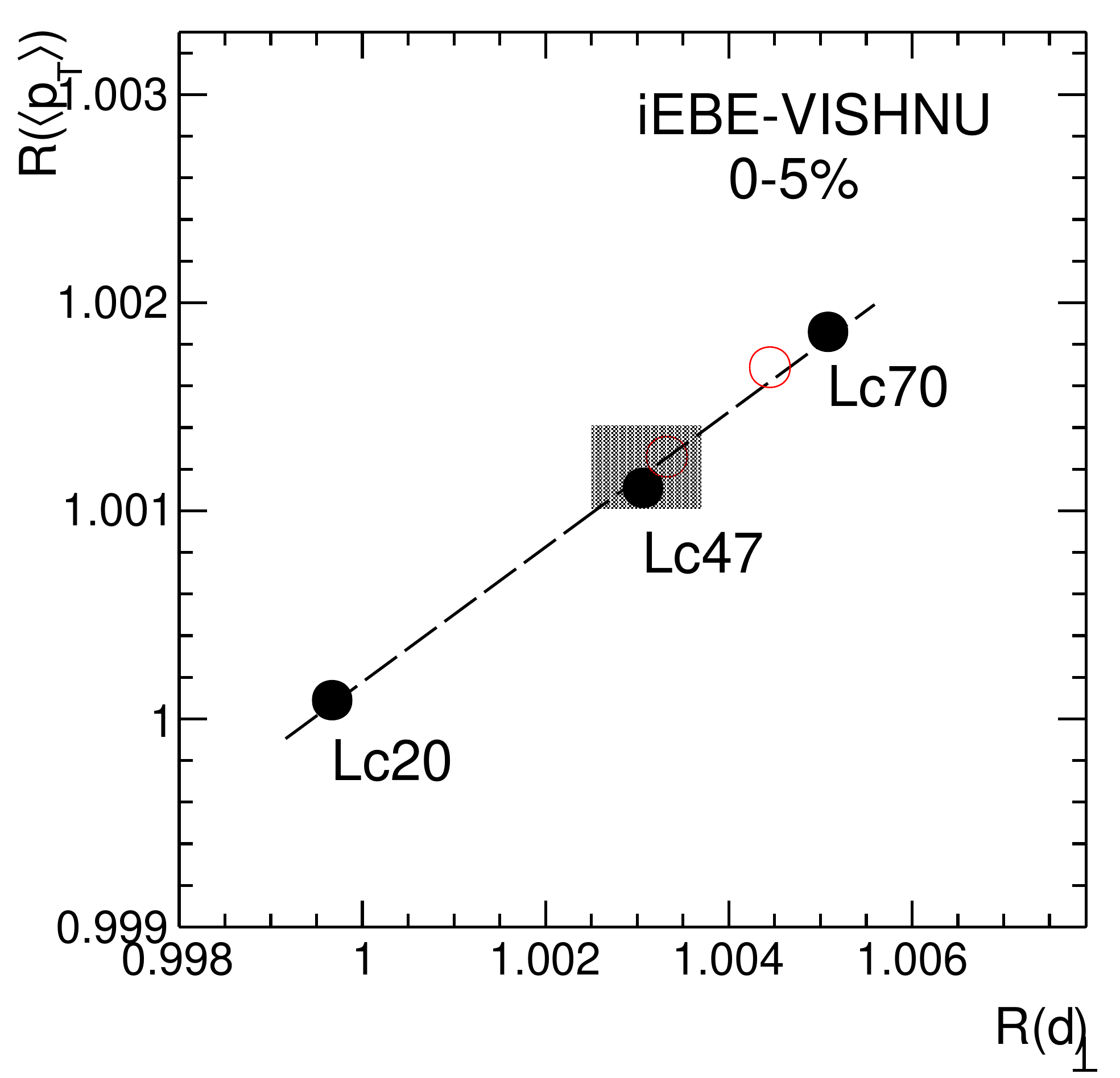}
      \caption{(Color online).
	  The \RuZr\ ratio $\Rpt$ in top $5\%$ centrality as a function of the transverse density ratio $\Rdt$. The filled markers show the results for spherical nuclear densities (Lc20, Lc47, and Lc70); the open circles show those for deformed Ru ($\beta_{2}=0.08$ and $0.16$) and spherical Zr corresponding to Lc47. The gray box indicates the uncertainties in $\Rpt$ arising from a factor of two change in both directions in the shear and bulk viscosities.
	  \label{fig:TopCentraltiy}} 
 \end{figure}

Clearly, $\Rpt$ depends on both the neutron skin thickness and nuclear deformation magnitude. 
We present in Fig.~\ref{fig:TopCentraltiy} the $\Rpt$ as a function of the \RuZr\ ratio of $\dt$, i.e.~$\Rdt=\dt^{\rm Ru+Ru}/\dt^{\rm Zr+Zr}$, for the three spherical densities in filled markers and for deformed densities corresponding to the Lc47 set in open circles. The deformation effectively reduces the size of the overlap area as indicated by Eq.~(\ref{eq:St}) where $\rt$ is matched to the spherical Lc47 density. This increases $\dt$ as seen in Fig.~\ref{fig:TopCentraltiy}.
An approximately linear relationship is observed between $\Rpt$ and $\Rdt$.
This confirms that $\meanpT$ is primarily dependent of $\dt$, which is in turn affected by the neutron skin and the deformation of the isobar nuclei.
We also show in Fig.~\ref{fig:TopCentraltiy} by the shaded box the effect of the factor of two variations in the shear and bulk viscosities in both directions; the effect is relatively small.

\begin{figure*}
	\includegraphics[scale=0.35]{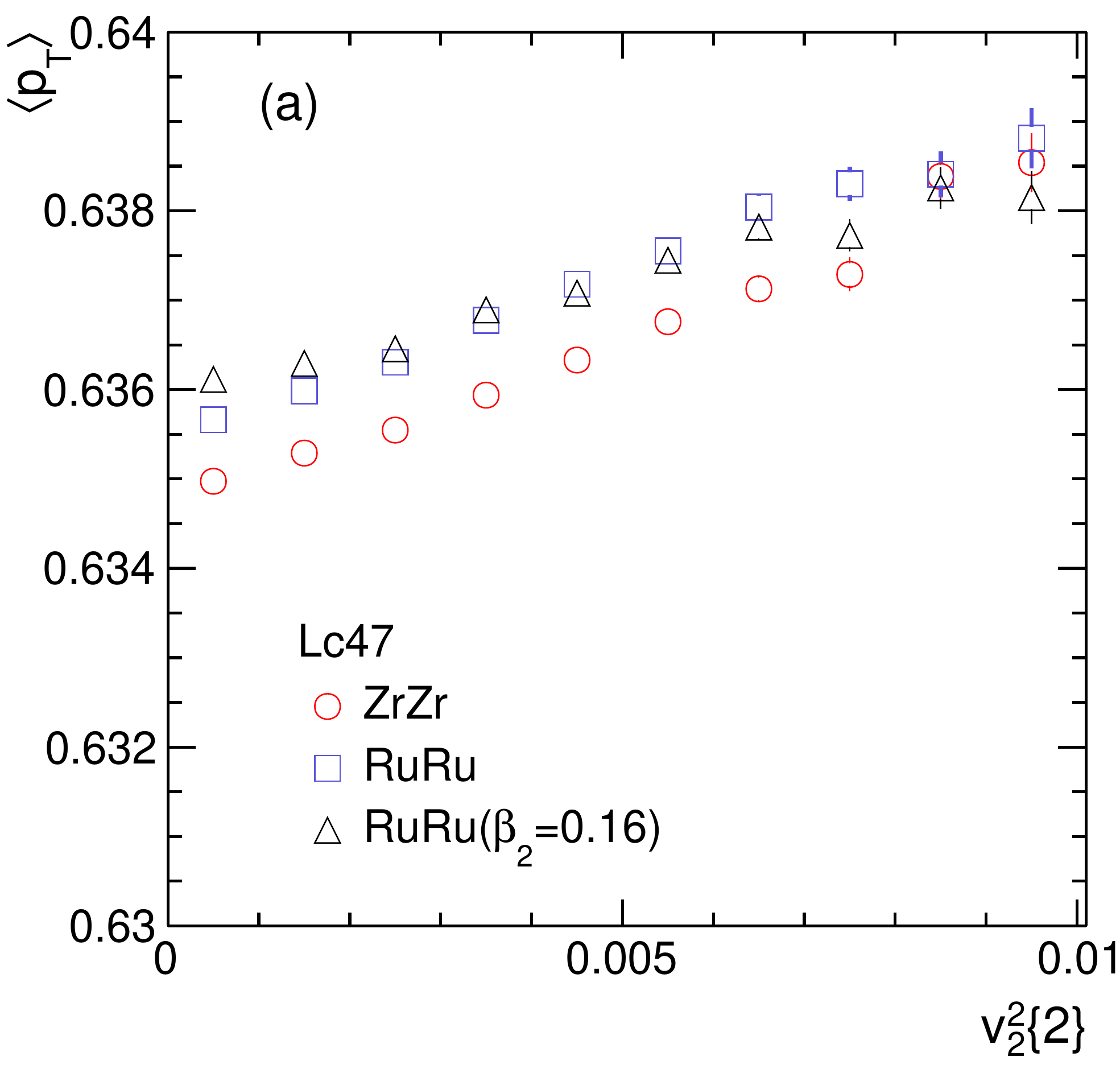}
	\includegraphics[scale=0.35]{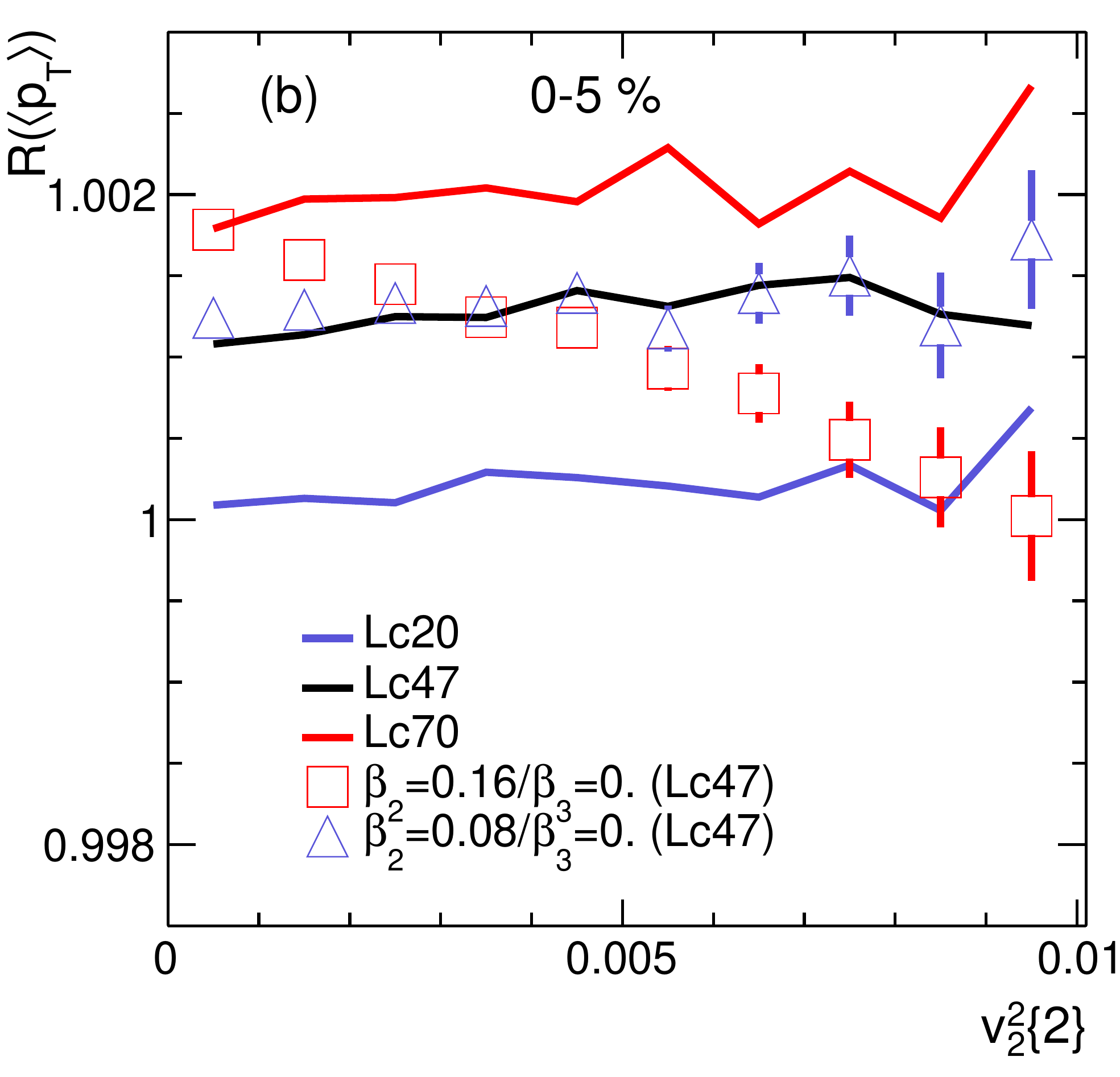}
      \caption{(Color online). (a) $\meanpT$ in spherical isobar collisions and in deformed Ru+Ru collisions, and (b) the \RuZr\ ratio $\Rpt$ as functions of the event-by-event $v_2^2\{2\}$ in the top 5\% centrality, calculated by the iEBE-VISHNU model. The curves in (b) are for spherical nuclei, and the data points are for the cases of deformed Ru and spherical Zr.}
      \label{fig:correl} 
 \end{figure*}

Both the neutron skin and deformation affect $\St$; they cannot be uniquely determined by a measurement of $\Rpt$.
However, nuclear deformations have large impact on anisotropic flow, particularly in central collisions~\cite{Heinz:2004ir,Filip:2009zz}. 
The conversion efficiency from the initial geometry anisotropy into the final-state momentum anisotropy depends on the collision dynamics and has strong model dependence. 
One may avoid those model dependence by exploiting the $v_n$ ratios in central collisions between the two isobar systems; one complication may be nonflow contamination in $v_n$ measurements. 
One may also resort to correlation measurement between $\meanpT$ and elliptic flow $v_2$, which has been widely discussed~\cite{Giacalone:2020dln,Giacalone:2020awm,Schenke:2020uqq}.
An anticorrelation between $\meanpT$ and $v_{2}$ has been found in central collisions with deformed U+U collisions~\cite{Giacalone:2020dln}. 
This arises from the positive correlation between $\mean{r_{\perp}^{2}}$ and geometry anisotropy, e.g., a tip-tip collision gives smaller $\mean{r_{\perp}^{2}}$ and $\ecc$ (and $v_{2}$). This positive correlation exceeds the anticorrelation caused by the $\sqrt{1-\ecc^2}$ term in Eq.~(\ref{eq:St}).
We found, however, from our hydrodynamic calculations that $\meanpT$ is positively correlated with $\ecc$ in all the studied cases, even for the Ru+Ru collisions with quadrupole deformation $\beta_{2}=0.16$; see Fig.~\ref{fig:correl}(a).
This is because the large 
fluctuations in the small isobar systems (compared to U+U) cause a significantly larger $\ecc$ in central collisions such that $\St$ is always anti-correlated with $\ecc$.

In the $\meanpT$ ratio between two isobar systems, however, the effects from fluctuations are largely cancelled and the difference in nuclear deformations survives.
This is shown in Fig.~\ref{fig:correl}(b), where $\Rpt$ is calculated, within the top 5\% centrality, in bins of $v_{2}^{2}$ which is computed by two-particle cumulant. An anticorrelation is observed between $\Rpt$ and 
$v_{2}^{2}$, the strength of which depends on the $\beta_2$ value. No such correlation is observed for the spherical nuclear densities.
This anticorrelation can be used to determine the quadrupole deformation difference between \Ru\ and \Zr, when the deformation is relatively large. 
Such determination can be relatively precise as it is insensitive to the nuclear densities, and may be immune to nonflow contamination in $v_2$.
For small deformation, it may be challenging to determine its magnitude, but its effect on $\Rdt$ is also small as shown in Fig.~\ref{fig:TopCentraltiy}.
Once the relative nuclear deformation is determined, the neutron skin thickness can be extracted from $\Rpt$.

{\em Conclusions.}
We have calculated the mean transverse momentum ratio $\Rpt$ between \RuRu\ and \ZrZr\ collisions with a (2+1)-dimensional viscous hydrodynamic model iEBE-VISHNU. The $\Rpt$ is found to be rather insensitive to the bulk properties of the collision systems, but remain sensitive to the small differences in the nuclear structure between the \Ru\ and \Zr\ nuclei.
Both the neutron skin thickness and the deformation affect the transverse overlap area $\St$ which primarily determines the $\meanpT$. It is found that the deformation can be determined from the correlation between $\Rpt$ and the elliptic flow $v_2$ in central isobar collisions. The neutron skin thickness can in turn be determined from $\Rpt$, which would complement low-energy nuclear interaction experiments to probe the symmetry energy density slope parameter.

\bibliography{ref}
\end{document}